\documentclass[aps,prd,preprint,superscriptaddress,tightenlines,nofootinbib]{revtex4}
\usepackage{epsfig}
\usepackage{graphicx}
\usepackage{dcolumn}
\usepackage{bm}

\newcommand{\etal}{{\it et al.}}

\begin{document}

\preprint{\tighten\vbox{\hbox{\hfil CLNS 05/1932}
                        \hbox{\hfil CLEO 05-20}}}


\title
{\LARGE Improved Measurement of ${\cal{B}}(D^+\to\mu^+\nu)$ and the
Pseudoscalar Decay Constant $f_{D^+}$}







\author{M.~Artuso}
\author{C.~Boulahouache}
\author{S.~Blusk}
\author{J.~Butt}
\author{O.~Dorjkhaidav}
\author{J.~Li}
\author{N.~Menaa}
\author{R.~Mountain}
\author{R.~Nandakumar}
\author{K.~Randrianarivony}
\author{R.~Redjimi}
\author{R.~Sia}
\author{T.~Skwarnicki}
\author{S.~Stone}
\author{J.~C.~Wang}
\author{K.~Zhang}
\affiliation{Syracuse University, Syracuse, New York 13244}
% ...
\collaboration{CLEO Collaboration} 
\noaffiliation

\date{August 29, 2005}

\begin{abstract}
We extract a relatively precise value for the decay constant of
the $D^+$ meson by measuring ${\cal{B}}(D^+\to\mu^+\nu)=(4.40\pm
0.66^{+0.09}_{-0.12})\times 10^{-4}$ using 281 pb$^{-1}$ of data
taken on the $\psi(3770)$ resonance with the CLEO-c detector. We
find $f_{D^+}=(222.6\pm 16.7^{+2.8}_{-3.4})~{\rm MeV},$ and
compare with current theoretical calculations.  We also set a 90\%
confidence upper limit on ${\cal{B}}(D^+\to e^+\nu)<2.4\times
10^{-5}$ which constrains new physics models.
\end{abstract}


\maketitle

Purely leptonic pseudoscalar meson decays proceed via the
annihilation of the constituent quarks into a virtual $W$ via the
axial-vector current. The decay rate is proportional to the square
of the decay constant $f$, a single number which encapsulates strong
interaction dynamics in the decay. The lack of hadrons in the final
state allows for precision tests of strong interaction theories.
Knowledge of decay constants is critical for extracting fundamental
information about CKM matrix elements. For example, $f_B$ is needed
to use measurements of $B\bar{B}$ mixing. Currently, it is not
possible to determine $f_B$ experimentally, so theoretical
calculations must be used. The most promising of these calculations
involves lattice QCD \cite{Davies}, though there are other methods.
Measurements of pseudoscalar decay constants from charm meson decays
provide checks on these calculations and can discriminate among
different models.


The decay $D^+\to \ell^+\nu$ proceeds by the $c$ and $\overline{d}$
quarks annihilating into a virtual $W^+$, with a decay width
\cite{Formula1}
\begin{equation}
\Gamma(D^+\to \ell^+\nu) = {{G_F^2}\over 8\pi}f_{D^+}^2m_{\ell}^2M_{D^+}
\left(1-{m_{\ell}^2\over M_{D^+}^2}\right)^2 \left|V_{cd}\right|^2~~~,
\label{eq:equ_rate}
\end{equation}
where $M_{D^+}$ is the $D^+$ mass, $m_{\ell}$ is the mass of the
final state lepton, $|V_{cd}|$ is a CKM matrix element that we
assume to be equal to $|V_{us}|$, and $G_F$ is the Fermi coupling
constant. Because of helicity suppression, the rate is a function of
$m_\ell^2$; consequently, the electron mode $D^+ \to e^+\nu$ has a
very small rate in the standard model. The relative widths are
$2.65:1:2.3\times 10^{-5}$ for the $\tau^+ \nu$, $\mu^+ \nu$, and
$e^+ \nu$ final states, respectively.

The CLEO-c detector is equipped to measure the momenta and
directions of charged particles, identify them using specific
ionization (dE/dx) and Cherenkov light (RICH), detect photons and
determine their directions and energies \cite{CLEODR}.


In this study we use 281 pb$^{-1}$ of data produced in $e^+e^-$
collisions using the Cornell Electron Storage Ring (CESR) and
recorded at the $\psi''$ resonance (3.770 GeV). This work contains
our previous sample as a subset and supersedes our initial effort
\cite{CLEODptomunu}.
 At this energy, the events consist mostly of
$D^+D^-$, $D^0\overline{D}^0$, three-flavor continuum $q\bar q$
($q=u,d,s$), $\tau^+\tau^-$, and $\gamma\psi'$ events.



Our analysis strategy is to fully reconstruct the $D^-$ meson in one
of six decay modes listed in Table~\ref{tab:Dreconnew} and search
for a $D^+\to \mu^+\nu$ decay in the rest of the event. Charge
conjugate modes are implicitly included throughout the paper. Track
selection, particle identification (PID), $\pi^0$, $K_S$, and muon
selection cuts are identical to those used in Ref.
\cite{CLEODptomunu}. We first evaluate $\Delta E$, the difference in
the energy of the decay products with the beam energy. The $\Delta
E$ distributions in all modes are well described by either a
Gaussian or the sum of two Gaussians, with root mean square (r.m.s.)
widths varying from 7 MeV for $D^- \to K^+K^-\pi^-$ to 14 MeV for
$D^-\to K^+ \pi^-\pi^-\pi^0$.  We select candidates by requiring
$|\Delta E|<$0.012 -- 0.024 GeV, where the cut in each mode is
approximately 2.5 times the r.m.s. width.

\begin{figure}[htb]
\centerline{\epsfxsize=3.0in \epsffile{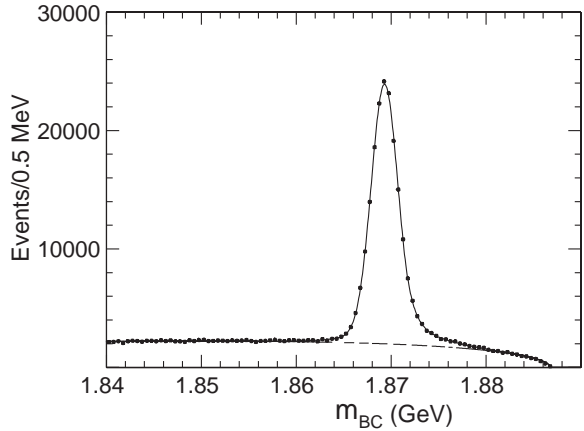}}
\caption{Beam-constrained mass for the sum of fully reconstructed
$D^-$ decay candidates. The solid curve shows the fit to the sum of
signal and background functions, while the dashed curve indicates
the background.} \label{Dreconnew}
\end{figure}

For the selected events we construct the beam-constrained mass,
$m_{BC}=\sqrt{E_{\rm beam}^2-(\sum_i \textit{\textbf{p}}_{i})^2},$
where $i$ runs over the final state particles from the candidate
$D^-$ decay. The resolution in $m_{BC}$ of 2.2-2.4 MeV is better
than merely calculating the invariant mass of the decay products,
since the CESR beam has a small energy spread. The $m_{BC}$
distribution for the sum of all $D^-$ tagging modes is shown in
Fig.~\ref{Dreconnew}. The numbers of tags in each mode are
determined from fits of the $m_{BC}$ distributions to a signal
function plus a background shape. For the latter we use an
expression analogous to one first used by the ARGUS collaboration to
approximate the correct threshold behavior \cite{ARGUS}.
For the signal we use an asymmetric lineshape because of the tail
towards high mass caused by initial state radiation \cite{CBL}.
Table~\ref{tab:Dreconnew} gives the numbers of signal and background
events for each mode within the signal region, defined as $m_D -
2.5\: \sigma_{m_{BC}}  < m_{BC} < m_D + 2.0\: \sigma_{m_{BC}}$,
where $\sigma_{m_{BC}}$ is the r.m.s. width of the lower side of the
distribution.

\begin{table}[htb]
\begin{center}
\begin{tabular}{lrcrc}\hline\hline
    Mode  &  \multicolumn{3}{c}{Signal}           &  Background \\ \hline
$K^+\pi^-\pi^- $ & 77387& $\pm$& 281   & $~1868$\\
$K^+\pi^-\pi^- \pi^0$ & 24850 &$\pm$& 214  & $12825$\\
$K_S\pi^-$ &   11162&$\pm$& 136& ~~~514\\
$K_S\pi^-\pi^-\pi^+ $ &  18176 &$\pm$& 255 & $~8976$\\
$K_S\pi^-\pi^0 $ &  20244&$\pm$& 170 & ~5223\\
$K^+K^-\pi^-$ & 6535&$\pm$& 95 &~1271 \\\hline
Sum &   158354&$\pm$& 496 & 30677\\
\hline\hline
\end{tabular}
\end{center}
\caption{Tagging modes and numbers of signal and background
events.
\label{tab:Dreconnew}}
\end{table}

Using our sample of $D^-$ candidates we search for events with a
single additional charged track presumed to be a $\mu^+$. The track
must make an angle $>$35.9$^{\circ}$ with respect to the beam-line,
deposit less than 300 MeV of energy in the calorimeter,
characteristic of a minimum ionizing particle, and not be
identified as a kaon.
We then calculate
\begin{equation}
{\rm MM}^2=\left(E_{\rm
beam}-E_{\mu^+}\right)^2-\left(-\textit{\textbf{p}}_{D^-}
-\textit{\textbf{p}}_{\mu^+}\right)^2, \label{eq:MMsq}
\end{equation}
where $\textit{\textbf{p}}_{D^-}$ is the three-momentum of the
fully reconstructed $D^-$. Real $D^+\to\mu^+\nu$ events will
congregate near zero MM$^2$.

The MM$^2$ from Monte Carlo simulation has a resolution
(two-Gaussian $\sigma$) of 0.0235$\pm$0.0004 GeV$^2$ consistent
among all the tagging decay modes. We check our simulations by using
the $D^+\to K_S\pi^+$ decay. Here
 we choose events with the same requirements as used to search for
 $\mu^+\nu$ but require one additional found $K_S$. The
 resolution is measured to be 0.0233$\pm$0.0009 GeV$^2$, consistent with the Monte
Carlo estimate.

In order to restrict the sample to candidate $\mu^+ \nu$ events, we
impose restrictions on tracks and neutral energy clusters in
addition to those used to reconstruct the tagging $D^-$. We select
events with only one additional charged track; events with extra
tracks originating within 0.5 m (radially) of the event vertex are
rejected. In addition, we eliminate events having at least one extra
neutral energy cluster of more than 250 MeV. These cuts are highly
effective in reducing backgrounds especially from $D^+\to
\pi^+\pi^0$ decays, but they introduce an inefficiency because the
decay products of the tagging $D^-$ can interact in the detector
material leaving spurious tracks or clusters. To evaluate our cut
efficiencies, we use an essentially background-free sample of fully
reconstructed $D^+D^-$ events. (The method is different here than in
our original publication, though the results are consistent.)

To first order, the fully reconstructed $D^-D^+\to K^+\pi^-\pi^-$,
$K^-\pi^+\pi^+$ events can be considered the superposition of two
 $D^-D^+\to K^+\pi^-\pi^-$, $\mu^+\nu$ events.
Our procedure is to evaluate the cut efficiency in our sample of
1435 events and take the square-root. This gives us the efficiency
for the $D^-\to K^+\pi^-\pi^-$ tag sample. We then combine the
$K^+\pi^-\pi^-$ with each of the other tags in turn. This method
ensures that the number of interactions of particles with material
and their resulting effects is the same as in the tag sample used
for the $\mu^+\nu$ analysis. In the sample of fully reconstructed
$D^-D^+$ events there are no events with extra tracks originating
within 0.5 m of the main event vertex. The efficiency for rejecting
events with extra clusters above 250 MeV, averaging over all our tag
modes, is (96.1$\pm$0.3$\pm$0.4)\% The systematic error arises only
because we have analyzed a situation corresponding to two
overlapping tags rather than one tag plus a muon. In Monte-Carlo
simulation the efficiency difference is 0.4\%, which we assign as a
systematic error.

\begin{figure}[htb]
\centerline{ \epsfxsize=3.0in \epsffile{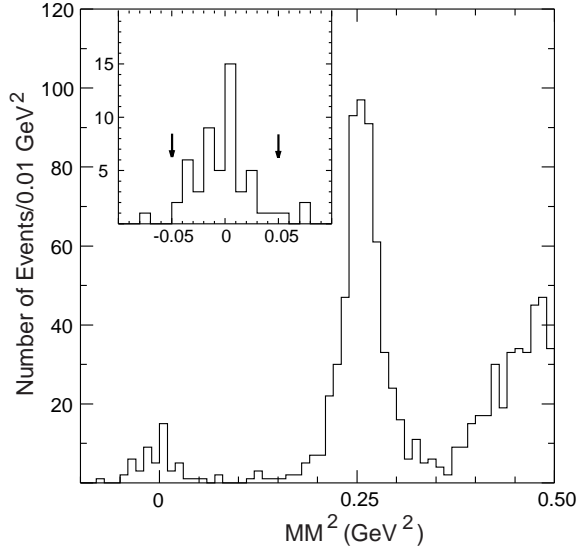} }
\caption{MM$^2$ using $D^-$ tags and one additional opposite sign
charged track and no extra energetic clusters (see text). The
insert shows the signal region for $D^+\to\mu^+\nu$ enlarged; the
defined signal region is shown between the two arrows.}
\label{mm2}
\end{figure}

The MM$^2$ distribution is shown in Fig.~\ref{mm2}.  We see a peak
near zero containing 50 events within the interval $-0.050$ GeV$^2$
to +0.050 GeV$^2$, approximately $\pm 2\sigma$ wide. The peak is
mostly due to $D^+\to\mu^+\nu$ signal. The large peak centered near
0.25 GeV$^2$ is from the decay $D^+\to \overline{K}^0\pi^+$ that is
far from our signal region and is expected, since many $K_L$ escape
our detector.

There are several potential background sources; these include other
$D^+$ modes, misidentified $D^0\overline{D}^0$ events, and continuum
including $e^+e^-\to\gamma\psi'$. Hadronic sources need to be
considered because the requirement of the muon depositing less than
300 MeV in the calorimeter, while about 99\% efficient on muons,
rejects only about 40\% of pions or kaons as determined from a pure
sample of $D^0\to K^{-}\pi^{+}$ decays.

There are a few specific $D^+$ decay modes that contribute
unwanted events in the signal region. Residual $\pi^+\pi^0$
background is determined from a simulation that uses a branching
fraction of (0.13$\pm$0.02)\% \cite{CLEOpipi} and yields
1.40$\pm$0.18$\pm$0.22 events; the first error is due to Monte
Carlo statistics, and the second is systematic, due mostly to the
branching ratio uncertainty. We find background from
$D^+\to\tau^+\nu$ only when $\tau^+\to \pi^+\nu$. Since the
$\tau^+\nu$ branching ratio is known to be 2.65 times the
$\mu^+\nu$ rate from Eq.~\ref{eq:equ_rate}, our simulation gives
1.08$\pm$0.15$\pm$0.16 events, where the systematic error arises
from our final uncertainty on the $\mu^+\nu$ decay rate. The
$\overline{K}^o\pi^+$ mode (branching ratio of (2.77$\pm$0.18)\%
\cite{PDG}) gives a large peak in the MM$^2$ spectrum near 0.25
GeV$^2$. While far from our signal region, the tail of the
distribution can contribute. We measure this background rate
directly using $D^0\overline{D^0}$ events. Here we select a sample
of single tags, either $K^-\pi^+\pi^+\pi^-$, $K^-\pi^+\pi^0$ or
$K^-\pi^+$ and look for events with only two additional oppositely
signed tracks, one identified as a kaon and one as a pion using
the RICH. We then compute the MM$^2$ ignoring the kaon. The MM$^2$
distribution shows a narrow peak near 0.25 GeV$^2$ and three
events in the signal region, corresponding to a background of
0.33$\pm$0.19$\pm$0.02 events, the systematic error being due to
the branching ratio uncertainty. (A simulation gives a consistent
estimate of 0.44$\pm$0.22 events.)

We have also checked the possibility of other $D^+D^-$ decay modes
producing background  with an equivalent 1.7 fb$^{-1}$ Monte Carlo
sample. We evaluate {$D^0\overline{D}^0$} and continuum
backgrounds by analyzing Monte Carlo samples corresponding to 0.54
fb$^{-1}$. To normalize our Monte Carlo events to our data sample,
we use $\sigma_{D^0\overline{D}^0}=3.5$ nb and $\sigma_{ \rm
continuum}=14.5$ nb \cite{sighad}. No additional background events
are found in any of these samples.

Our total background is 2.81$\pm$0.30$\pm$0.27 events. The
backgrounds from other $D^+$, $D^0$, and continuum sources are
limited to less than 0.4, 0.4, and 1.2 events at 90\% confidence
level (C.L.), respectively. To account for possible backgrounds from
these sources, we add them as 32\% C.L. (1$\sigma$) values in
quadrature for a positive error and therefore add an additional
$^{+0.8}_{-0}$ event systematic error.



We have 47.2$\pm 7.1^{+0.3}_{-0.8}$ $\mu^+\nu$ signal events after
subtracting background. The detection efficiency for the single muon
of 69.4\% includes the selection on MM$^2$ within $\pm2\sigma$
limits, the tracking, the particle identification, probability of
the crystal energy being less than 300 MeV, and corrections for
final state radiation \cite{gammamunu}. It does not include the
96.1\% efficiency of not having another unmatched cluster in the
event with energy greater than 250 MeV. We also need to account for
the fact that it is easier to find tags in $\mu^+\nu$ events than in
generic decays by a small amount, (1.5$\pm$0.4$\pm$0.5)\%, as
determined by Monte Carlo simulation.

Our result for the branching fraction, using the tag sum in
Table~\ref{tab:Dreconnew}, is
\begin{equation}
{\cal{B}}(D^+\to\mu^+\nu)=(4.40\pm 0.66^{+0.09}_{-0.12})\times
10^{-4}~.
\end{equation}
The systematic errors on the branching ratio are listed in
Table~\ref{tab:eff}. (The systematic error on the tag sum is
estimated from varying the signal and background functions.)
\begin{table}[htb]
\begin{center}
\begin{tabular}{lc}\hline\hline
     &Systematic errors (\%) \\ \hline
MC statistics &$\pm$0.4  \\
Track finding &$\pm$0.7 \\
PID cut &$\pm$1.0 \\
MM$^2$ width & $\pm$1.0\\
Minimum ionization cut &$\pm$1.0 \\
Number of tags& $\pm$0.6\\
Extra showers cut & $\pm$0.5 \\
Background  & ${+0.6}$, ${-1.7}$\\\hline
Total &${+2.1}$, ${-2.5}$\\
 \hline\hline
\end{tabular}
\end{center}
\caption{Systematic errors on the $D^+ \to \mu^+ \nu$ branching
ratio. \label{tab:eff}}
\end{table}

The decay constant $f_{D^+}$ is then obtained from
Eq.~(\ref{eq:equ_rate}) using 1.040$\pm$0.007 ps as the $D^+$
lifetime \cite{PDG}, and $|V_{cd}|$ = 0.2238$\pm$0.0029
\cite{KTeV}. (We add these two small additional sources of
uncertainty into the systematic error.) Our final result is
\begin{equation}
f_{D^+}=(222.6\pm 16.7^{+2.8}_{-3.4})~{\rm MeV}~.
\end{equation}


We use the same tag sample to search for $D^+\to e^+\nu_{e}$. We
identify the electron using a match between the momentum
measurement in the tracking system and the energy deposited in the
CsI calorimeter as well as insuring that the shape of the energy
distribution among the crystals is consistent with that expected
for an electromagnetic shower. Other cuts remain the same. We do
not find any candidates, yielding a 90\% C.L. limit of
${\cal{B}}(D^+\to e^+\nu_{e})<2.4\times 10^{-5}~,$ including
systematic errors.


Our measurement of $f_{D^+}$ is much more precise than previous
observations or limits \cite{CLEODptomunu,MarkIII}. The theoretical
predictions listed in Table~\ref{tab:Models} were made prior to this
result. The first entry is the result from the Fermilab-MILC-HPQCD
collaboration that is done with all three light quark flavors
unquenched, hence $n_f$=2+1 \cite{Lat:Milc}. It is about 10\%
smaller than our result, albeit within error.

\begin{table}[htb]
\begin{center}
\begin{tabular}{lcl}\hline\hline
    Model  &  $f_{D^+}$ (MeV)          &  ~~~~~$f_{D_S^+}/f_{D^+}$           \\\hline
Lattice ($n_f$=2+1)  \cite{Lat:Milc} &
$201\pm 3 \pm 17 $&$1.24\pm 0.01\pm 0.07$ \\
QL (Taiwan) \cite{Lat:Taiwan} &
$235 \pm 8\pm 14 $&$1.13\pm 0.03\pm 0.05$ \\
QL (UKQCD) \cite{Lat:UKQCD} & $210\pm 10^{+17}_{-16}$ & $1.13\pm 0.02^{+0.04}_{-0.02}$\\
QL \cite{Lat:Damir} & $211\pm 14^{+0}_{-12}$ &
$1.10\pm 0.02$\\
QCD Sum Rules \cite{Chiral} & $203\pm 20$ & $1.15\pm 0.04$ \\
QCD Sum Rules \cite{Sumrules} & $195\pm 20$ & \\
Quark Model \cite{Quarkmodel} & $243\pm 25$ & 1.10 \\
Potential Model \cite{Equations} & 238  & 1.01 \\
Isospin Splittings \cite{Isospin} & $262\pm 29$ & \\
\hline\hline
\end{tabular}
\end{center}
\caption{Theoretical predictions of $f_{D^+}$ and
$f_{D_S^+}/f_{D^+}$. QL indicates quenched lattice calculations.}
\label{tab:Models}
\end{table}

The models generally predict $f_{D_S^+}$ to be 10--25\% larger than
$f_{D^+}$ which is consistent with a previous CLEO measurement
\cite{chadha}. Some non-standard models predict significant rates
for the helicity suppressed decay $D^+\to e^+\nu$ \cite{Akeroyd}.
Our upper limit restricts these models.

We gratefully acknowledge the effort of the CESR staff in providing
us with excellent luminosity and running conditions. This work was
supported by the National Science Foundation and the U.S. Department
of Energy.

\newpage

\end{document}